\documentclass[a4paper,twocolumn,american,prl,superscriptaddress,longbibliography,fixfloat,notitlepage]{revtex4-1}
\usepackage[latin9]{inputenc}
\usepackage{amsmath}
\usepackage{amssymb}
\usepackage{graphicx}

\makeatletter

\pdfpageheight\paperheight
\pdfpagewidth\paperwidth

\usepackage{xcolor}
\definecolor{darkblue}{rgb}{0.1,0.2,0.6} 
\definecolor{lightblue}{rgb}{0.1,0.1,1.0}
\definecolor{darkred}{rgb}{0.8,0.1,0.2}
\usepackage[colorlinks,citecolor=lightblue,linkcolor=darkblue,urlcolor=lightblue] {hyperref}

\makeatother

\usepackage{babel}
\begin{document}
\title{Suppression of heating by long-range interactions in periodically
driven spin chains}
\author{Devendra Singh Bhakuni}
\affiliation{Department of Physics, Ben-Gurion University of the Negev, Beer-Sheva
84105, Israel}
\author{Lea F. Santos}
\affiliation{Department of Physics, Yeshiva University, New York, New York 10016,
USA}
\author{Yevgeny Bar Lev}
\affiliation{Department of Physics, Ben-Gurion University of the Negev, Beer-Sheva
84105, Israel}
\email{ybarlev@bgu.ac.il}

\begin{abstract}
We propose a mechanism to suppress heating in periodically driven
many-body quantum systems by employing sufficiently long-range interactions
and experimentally relevant initial conditions. The mechanism is robust
to local perturbations and does \emph{not} rely on disorder or high
driving frequencies. Instead, it makes use of an approximate fragmentation
of the many-body spectrum of the non-driven system into bands, with
band gaps that grow with the system size. We show that when these
systems are driven, there is a regime where \emph{decreasing} the
driving frequency \emph{decreases} heating and entanglement build-up.
This is demonstrated numerically for a prototypical system of spins
in one dimension, but the results can be readily generalized to higher
dimensions.
\end{abstract}
\maketitle
Periodically driven quantum systems continue to produce fascinating
physics and phenomena inaccessible to their static counterparts. Some
notable examples include the Kapitza pendulum~\citep{kapitza1965dynamical},
dynamical localization~\citep{dunlap1986dynamic,dunlap1988dynamic,eckardt2009exploring},
Floquet topological insulators~\citep{cayssol2013floquet,gomez-leon2013floquet,rudner2013anomalous},
dynamical phase transitions~\citep{bastidas2012nonequilibrium},
induced many-body localization (MBL)~\citep{choi2018dynamically,dalessio2013many,bairey2017driving,bhakuni2020drive,lazarides2015fate},
and Floquet time-crystals~\citep{else2016floquet,khemani2016phase,yao2017discrete,zhang2017observation,kshetrimayum2020stark}.
However, a key obstacle to realizing new phases of matter in driven
systems is that typically the drive heats up the system to a featureless
infinite-temperature state where all correlations and observables
become trivial~\citep{dalessio2014long,ponte2015periodically,luitz2017absence,lazarides2014equilibrium}.

In one-dimensional systems, heating can be suppressed with the inclusion
of sufficiently strong disorder, which leads to the formation of the
Floquet-MBL phase~\citep{ponte2015many,lazarides2015fate,ponte2015periodically,abanin2016theory,bordia2017periodically}.
Alternatively, heating can be suppressed at any dimension, whether
the system is clean or disordered, by considering driving frequencies
greater than the single-particle excitation energy, such that the
absorption of a photon from the drive will always result in a multiparticle
process \citep{abanin2015exponentially,abanin2017effective,mori2016rigorous,abanin2017rigorous,weidinger2017floquet,singh2019quantifying,santos2021}.
Under these conditions the system will spend a significant amount
of time in a nontrivial metastable state -- a phenomenon called Floquet-prethermalization~\citep{abanin2015exponentially,abanin2017effective,mori2016rigorous,abanin2017rigorous,weidinger2017floquet,singh2019quantifying,santos2021}.
It has been recently demonstrated with nuclear spins using nuclear
magnetic resonance techniques~\citep{peng2021floquet} and with ultracold
atoms in a driven optical lattice~\citep{rubio-abadal2020floquet}.

If the driving frequencies are smaller than the single-particle excitation
energy, the system can efficiently absorb energy from the drive, which
results in fast heating to infinite temperature~\emph{}\footnote{\textit{Recently, Floquet-prethermalization was achieved away from
the high frequency limit in a model with short-range interactions
by imposing special constraints on the driving protocol }\citep{fleckenstein2021prethermalization}\textit{.}}. But is this the fate of all driven quantum systems?\emph{ }In this
Letter, we show that the answer is negative. Heating can actually
be suppressed in any dimension and for frequencies smaller than the
single-particle excitation if the system has sufficiently long-range
interactions.

The physics of non-driven systems with power-law decaying interactions,
$r^{-\alpha}$ (where $r$ is the distance between two bodies), has
gained considerable attention due to experimental realizations in
trapped ions~\citep{richerme2014nonlocal,jurcevic2014quasiparticle,gring2012relaxation,morong2021observation,kyprianidis2021observation},
where the range of the interactions can be tuned. A particularly intriguing
regime is $\alpha<d$ ($d$ being the dimension of the system), where
conventional thermodynamics does not apply~\citep{dauxois2002dynamics}.
Power-law decaying interactions occur in various systems, from spin
glasses and magnetically frustrated systems to atomic, molecular,
and optical systems~\citep{binder1986spin,saffman2010quantum,yan2013observation,islam2013emergence,britton2012engineered}.
They are associated with phenomena that are absent for neighboring
interactions~\citep{mermin1966absence,mukamel2005breaking,celardo2006time,bachelard2008abundance,kastner2010nonequivalence,schachenmayer2013entanglement,eisert2013breakdown,hauke2013spread}.
They are known to affect transport \citep{levitov1990delocalization,aleiner2011localization,gutman2016energy,prasad2021many,kloss2019spin,kloss2020spin,chavez2021disorder},
destroy many-body localization~\citep{burin2006energy,burin2015localization,deroeck2017stability,tikhonov2018many,gopalakrishnan2019instability,nag2019many,roy2019self},
and facilitate the propagation of correlations ~\citep{hauke2013spread,eisert2013breakdown,gong2014persistence,mazza2014out,foss-feig2015nearly}.

While for $\alpha>d$ , the physics is many times only quantitatively
different from the physics of systems with local interactions ($\alpha\to\infty$),
novel physics often emerges for slowly decaying interactions, $\alpha<d$.
An example is the emergence of a Hilbert space fragmentation into
weakly connected subspaces. If the dynamics starts in one of these
subspaces, it can be effectively described by a local Hamiltonian
for a long time~\citep{santos2016cooperative,celardo2016shielding},
so despite the presence of long-range interactions, features that
are usually associated with short-range interactions may be observed,
such as the logarithmic growth of entanglement~\citep{schachenmayer2013entanglement,lerose2020origin},
light-cone evolution~\citep{santos2016cooperative,storch2015interplay},
and self-trapping~\citep{nazareno1999long}. On the other hand, if
the initial state spans multiple subspaces, the dynamics violates
the generalized Lieb-Robinson bound and leads to the instantaneous
spread of correlations~\citep{richerme2014nonlocal,jurcevic2014quasiparticle,gring2012relaxation,hauke2013spread}.

The behavior of periodically driven systems with power-law decaying
interactions was studied in~\citep{ho2018bounds,machado2019exponentially,machado2020long,kuwahara2016floquet}.
For $\alpha>d$ and large driving frequencies, exponentially slow
heating and the emergence of Floquet prethermalization were obtained~\citep{abanin2015exponentially,abanin2017effective,mori2016rigorous,abanin2017rigorous,machado2019exponentially}.
In this prethermal regime, a novel non-equilibrium phase of matter
dubbed the prethermal time-crystal~\citep{machado2020long}, which
is similar to the MBL-time crystal~\citep{else2016floquet,khemani2016phase,yao2017discrete},
has been argued to exist. For $\alpha<d$, the general expectation
is that to achieve a prethermal plateau, the system needs both to
be in the \emph{high-frequency regime} and to be finite. The second
condition arises since the single-particle excitation energy increases
with system size and therefore for fixed frequency, the prethermal
plateau shrinks as the system size increases~\citep{machado2019exponentially}.

In this Letter, we show that it is in fact possible to suppress heating
in systems with long-range interactions in the \emph{low-frequency
regime}, where the driving frequencies are smaller than the single-particle
excitation energy. This can be done by taking advantage of the effective
fragmentation of the Hilbert space, which is induced by interactions
with $\alpha<d$, and by selecting initial states within one of those
approximate subspaces, such that the energy absorption from the drive
becomes ineffective. In this way, we can achieve prethermal phases
whose lifetimes grow as the system size increases and which are viable
at any dimension. We demonstrate this behavior by numerically examining
the dynamics of the half-chain entanglement entropy and the energy
absorption in a spin chain with $\alpha<1$.

\textit{Model}.---We consider a long-range interacting spin chain
of length $L$ described by the Hamiltonian, 
\begin{align}
\hat{H}_{0}= & J_{z}\hat{V}+J_{x}\sum_{\langle i,j\rangle}^{L-1}\hat{\sigma}_{i}^{x}\hat{\sigma}_{j}^{x}+h_{x}\sum_{i=1}^{L}\hat{\sigma}_{i}^{x};\\
\hat{V}= & \sum_{i<j}^{L-1}\frac{1}{\left|i-j\right|{}^{\alpha}}\hat{\sigma}_{i}^{z}\hat{\sigma}_{j}^{z},\nonumber 
\end{align}
where $\hat{\sigma}_{i}^{x,y,z}$ are Pauli operators, $J_{z}$ is
the strength of the long-range term $\hat{V}$ and we set $J_{z}=1$,
$J_{x}$ corresponds to the strength of nearest-neighbor interactions
in the $x$-direction, and $h_{x}$ is the amplitude of a transverse
magnetic field. The operator norm of the long-range term is $\left\Vert \hat{V}\right\Vert \sim L^{2-\alpha}$
for $\alpha<1$, such that it becomes dominant in the thermodynamic
limit. Nevertheless, even in this limit, the dynamics is \emph{not
}given by $\hat{V}$ for almost all initial states, and is highly
nontrivial, since the model stays non- integrable for any value of
$\alpha$. While one can make $\hat{V}$ extensive by proper rescaling
~\citep{kastner2011diverging,bachelard2013universal,kastner2017nscaling},
this rescaling does not naturally occur in the experiments, where
finite systems are studied~\citep{richerme2014nonlocal,jurcevic2014quasiparticle,neyenhuis2017observation,zhang2017observation}.
We therefore do not consider this rescaling in our work.

The static Hamiltonian $\hat{H}_{0}$ is periodically driven by the
following time-dependent perturbation, 
\begin{equation}
\hat{H}_{1}\left(t\right)=\text{sgn}\left(\sin\left(\omega t\right)\right)\left(h_{y}\sum_{i=1}^{L}\hat{\sigma}_{i}^{y}+h_{z}\sum_{i=1}^{L}\hat{\sigma}_{i}^{z}\right),\label{eq2}
\end{equation}
such that the total Hamiltonian is $\hat{H}\left(t\right)=\hat{H}_{0}+\hat{H}_{1}\left(t\right)$.
Here, $\omega=2\pi/T$ is the driving frequency, $\text{sgn}\left(.\right)$
is the sign function, $T$ is the driving period, and $h_{y}$ and
$h_{z}$ are the magnitudes of the magnetic fields along the $y$-
and $z$-directions, respectively. We use a square-wave driving to
closely follow the experiment with trapped ions in Ref.~\citep{kyprianidis2021observation},
and also since it is computationally more efficient than a continuous
time-varying drive. However, the results presented here should be
insensitive to the choice of the driving protocol. We explore the
dynamics of the driven system, $\hat{H}\left(t\right)$, with $\alpha<1$.

To study the heating dynamics, we use the numerically exact Krylov
subspace techniques to evolve the system in time~\citep{luitz2017ergodic}.
Due to the lack of symmetries, we have to consider the entire Hilbert
space of dimension $2^{L}$, so we analyze system sizes up to $L=22$.
We investigate the energy density of the static system measured with
respect to the initial state, 
\begin{equation}
\varepsilon\left(t\right)\equiv\frac{1}{L}\text{Tr }\left[\left(\hat{\rho}\left(0\right)-\hat{\rho}\left(t\right)\right)\hat{H}_{0}\right],
\end{equation}
where $\hat{\rho}\left(t\right)$ is the density matrix as a function
of time, and the half-chain entanglement entropy, 
\begin{equation}
S\left(t\right)=-\text{Tr}\left[\hat{\rho}_{A}\left(t\right)\text{ln}\hat{\rho}_{A}\left(t\right)\right],
\end{equation}
where $\hat{\rho}_{A}\left(t\right)=\text{Tr}_{B}\hat{\rho}\left(t\right)$
is the reduced density matrix of the subsystem $A$ consisting of
$L/2$ spins.

\begin{figure}
\includegraphics[width=8.6cm]{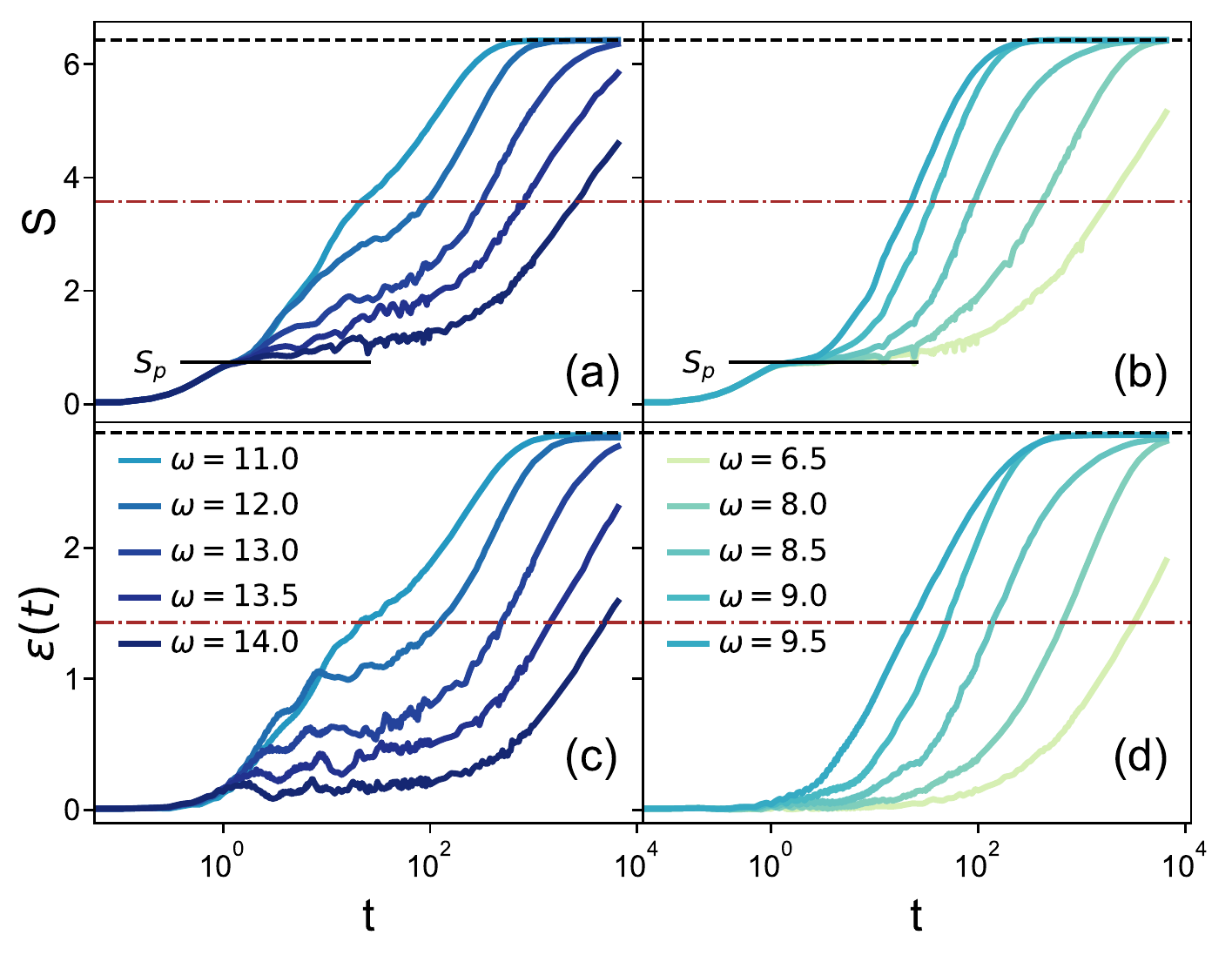} \caption{Dynamics of the half-chain entanglement entropy (a,b) and the energy
absorption (c,d) for different ranges of the driving frequencies $\omega$.
The infinite-temperature values are marked by horizontal black dashed
lines and the prethermal values of $S(t)$ by solid horizontal lines.
The dot-dashed red lines mark the heating time, where the entropy
(energy) reaches the half-way mark between its plateau value (initial
value) and its infinite-temperature value. The initial state is $|\psi(0)\rangle=|11\cdots11011\cdots11\rangle$,
$L=20,\alpha=0.67,J_{x}=0.69,h_{x}=0.23,h_{y}=0.21$, and $h_{z}=0.19$.
For these parameters, $J_{\text{eff}}=\Delta_{1}=10.92$.}
\label{Fig_1}
\end{figure}

\textit{Heating suppression}.--- Figure~\ref{Fig_1} shows the evolution
with time of the entanglement entropy and the energy density for $L=20$,
different frequencies, and the initial state $\left|\psi(0)\right\rangle =|11\cdots11011\cdots11\rangle$,
where all the spins, except the one in the middle, point up. For most
frequencies in Fig.~\ref{Fig_1}, the entanglement entropy exhibits
three distinct regimes: an initial growth for a short time, which
is followed by the emergence of a long-lived prethermal state (Floquet-prethermalization),
where \emph{$S(t)$ }saturates to a plateau value $S_{p}$ (horizontal
black solid line), after which the entropy finally reaches an infinite-temperature
value (black dashed line) corresponding to the result by Page, $S_{\text{Page}}=(L\ \text{ln}\,2-1)/2$~\citep{page1993average}.
The dependence of the behavior of the energy density on the frequency
is comparable to that for the entropy, it remains constant during
the prethermal phase and eventually goes to its infinite-temperature
value at long-times. Those distinct dynamical stages in Fig.~\ref{Fig_1}
were observed before in Ref.~\citep{machado2019exponentially}, where
high driving frequencies were considered and the dynamics started
with initial product states in the $z$-direction with an equivalent
number of spins pointing up and down. But in stark contrast with previous
studies, we find that \emph{below} a certain frequency value, we can
extend the prethermal phase and postpone heating by \emph{decreasing}
the driving frequency, as shown in Figs.~\ref{Fig_1}~(b,d). Contrary
to past studies for which the heating time increases monotonically
with the frequency, we have now a non-monotonic dependence. For frequencies
$\omega\gtrsim11$, the heating time grows as $\omega$ increases
{[}Fig.~\ref{Fig_1}~(a,c){]}, but for a range of frequencies with
$\omega<11$, the heating time shrinks as $\omega$ increases {[}Fig.~\ref{Fig_1}~(b,d){]}.
For frequencies close to $\omega\sim11$, the system heats up very
quickly, hinting on a resonant behavior.

\begin{figure}
\includegraphics[width=8.6cm]{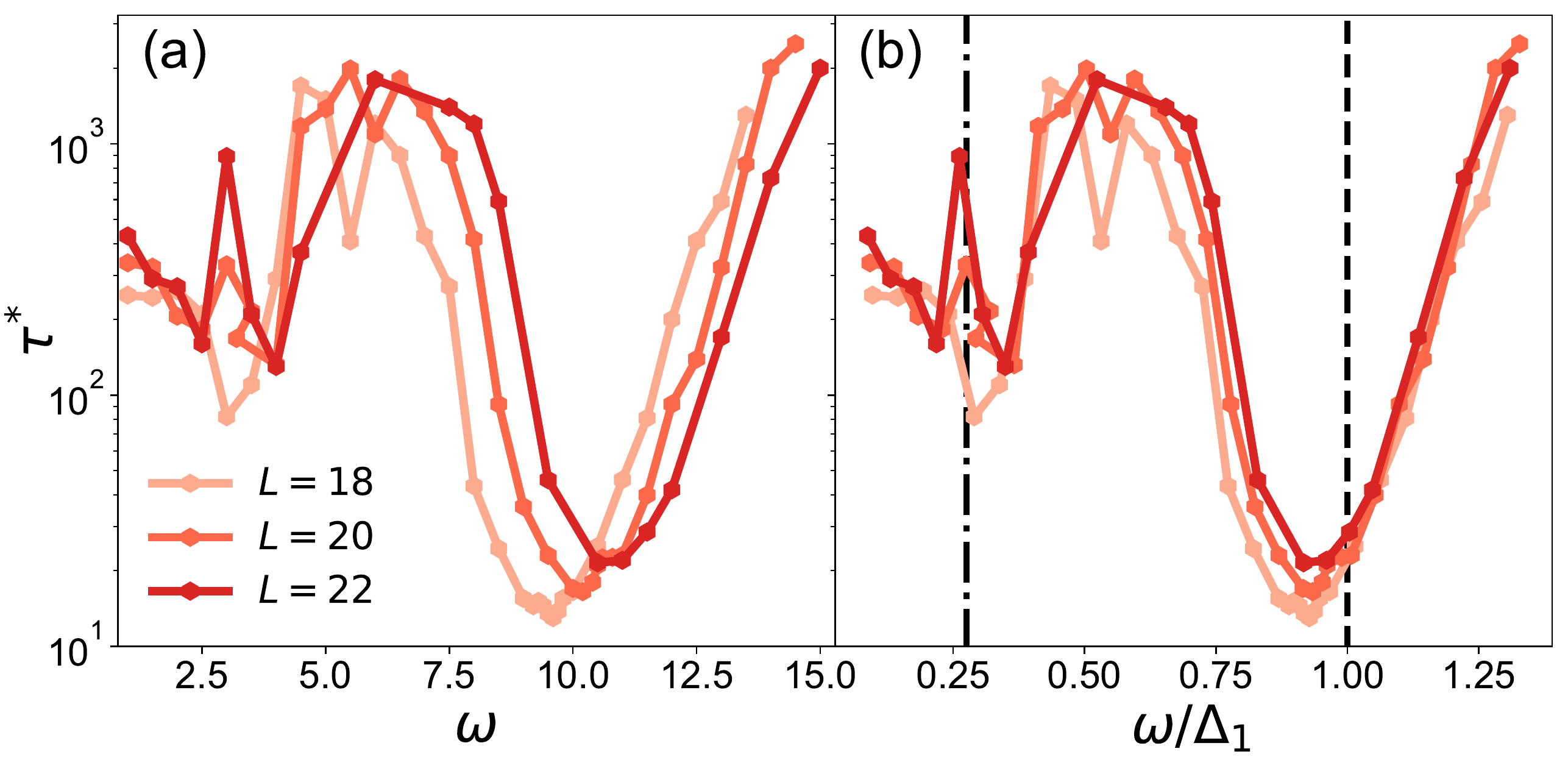}\caption{Heating time, $\tau^{*}$, as extracted from the entanglement entropy
for different system sizes as a function of (a) frequency and (b)
rescaled frequency, $\omega/\Delta_{1}$. The two vertical lines on
the x-axis in (b) indicate the gaps $\Delta_{2}$ (dashed-dot) and
$\Delta_{1}$ (dashed) for $L=22$. The gap $\Delta_{1}$ is equal
to $J_{\text{eff}}$ (see text). As in Fig.~\ref{Fig_1}: initial
state $|\psi(0)\rangle=|11\cdots11011\cdots11\rangle$, $\alpha=0.67,J_{x}=0.69,h_{x}=0.23,h_{y}=0.21$,
and $h_{z}=0.19$.}
\label{Fig_2}
\end{figure}

To show the frequency dependence more explicitly, we define the heating
time $\tau^{*}$ as the time when the entanglement entropy reaches
a half-way mark between its prethermal plateau and its asymptotic
value, $S(\tau^{*})\equiv S_{p}+\left[S_{\text{Page}}-S_{p}\right]/2$,
which is indicated with dot-dashed red lines in Figs.~\ref{Fig_1}~(a,b).
We see in Fig.~\ref{Fig_2}~(a) that, as expected, for $\omega>11$
the heating time increases as we increase the driving frequency, however,
within a range of values for $\omega<11$, the heating time increases
as we decrease $\omega$ and these results further improve as the
system size grows. A similar qualitative picture is obtained also
for the heating time calculated from the energy density $\varepsilon\left(t\right)$.
As we explain next, this unusual dependence on the frequency is a
consequence of the effective fragmentation of the Hilbert space verified
for the non-driven system when $\alpha<1$~\citep{santos2016cooperative}.

\textit{Energy bands}.--- To better understand the Hilbert space
fragmentation of the static system, let us first examine the long-range
term $\hat{V}$ of $\hat{H}_{0}$ {[}Eq.~(1){]}, which for $\alpha=0$
can be written in terms of the collective spin operator $\hat{M}_{z}=$$\sum_{i}^{L}\hat{\sigma}_{i}^{z}/2$
as $\hat{V}=$$2\hat{M}_{z}^{2}-L/2$. The energy spectrum of $\hat{V}$
consists of degenerate bands with the energies 
\begin{equation}
E_{b}=2\left(\frac{L}{2}-b\right)^{2}-\frac{L}{2},\qquad b=0,1,..,\frac{L}{2},
\end{equation}
where $b$ indicates the number of spins pointing down in the $z$-direction,
and we designate the corresponding energy band as the band $b$. Since
the energy of a product state with $b$ down-spins is equal to the
energy of a state with $L-b$ down-spins, each band is $2{L \choose b}$
degenerate for $b<L/2$. For $0<\alpha<1$, the degeneracy within
each band of the spectrum of $\hat{V}$ is partially lifted, but the
different subspaces are still separated in energy. We define the energy
gap between two nearby bands as $\Delta_{b}\equiv E_{b}-E_{b-1}$,
which can be obtained analytically. The gap between the bands $b=0$
and $b=1$ can be calculated as
\begin{equation}
\Delta_{1}=\sum_{r=1}^{L-1}\frac{2}{r^{\alpha}}\sim\frac{2}{1-\alpha}L^{1-\alpha}.\label{eq:delta1}
\end{equation}
One sees that the gap increases monotonically with system size for
$\alpha<1$. Similarly, we can obtain $\Delta_{2}$, 
\begin{align}
\Delta_{2} & =\left(\sum_{r=2}^{L-1}\frac{2}{r^{\alpha}}+\sum_{r=1}^{L-2}\frac{2}{r^{\alpha}}\right)-2\left(\sum_{r=1}^{L/2-1}\frac{2}{r^{\alpha}}+\left(\frac{2}{L}\right)^{\alpha}\right)\nonumber \\
 & \sim\frac{2\left(2-2^{\alpha}\right)}{1-\alpha}L^{1-\alpha},\label{eq:delta2}
\end{align}
which also increases with the system size, although $\Delta_{2}<\Delta_{1}$.

\begin{figure}
\includegraphics[width=7.6cm]{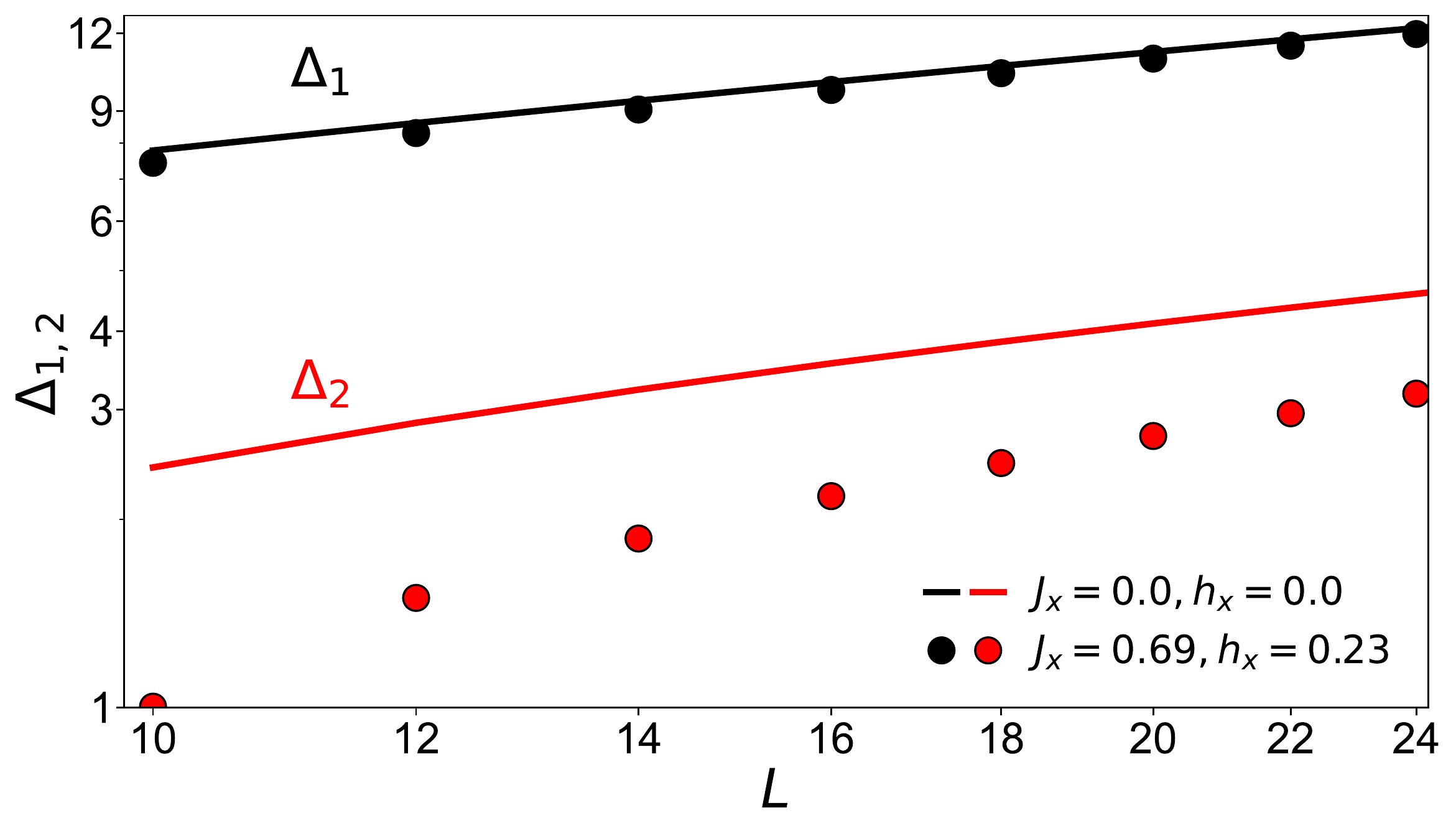}\caption{The lowest energy gaps $\Delta_{1}$ (black lines and circles) and
$\Delta_{2}$ (red lines and circles) as a function of $L$. The solid
lines indicate the exact calculations for $J_{x}=h_{x}=0$ {[}Eq.(\ref{eq:delta1})
and Eq.(\ref{eq:delta2}){]} and the circles are the numerical results
for $J_{x}=0.69$ and $h_{x}=0.23$.}
\label{Fig_3}
\end{figure}

The other terms of the static Hamiltonian $\hat{H}_{0}$ couple the
states of $\hat{V}$. The $J_{x}$ term connects states within the
same band and states from band $b$ to bands $b\pm2$, while the $h_{x}$
term connects states of band $b$ to bands $b\pm1$. However, if the
values of $J_{x}$ and $h_{x}$ are smaller than the gap between the
bands, they cannot effectively couple them. Furthermore, the numerical
calculations for the values of $\Delta_{1}$ and $\Delta_{2}$ for
$\hat{H}_{0}$ with $J_{x},h_{x}\neq0$ approach the gaps between
the bands of $\hat{V}$ in the limit $L\to\infty$, as shown in Fig.~\ref{Fig_3}.
This implies that the dynamics starting from an initial state within
one band gets confined to that approximate subspace for a time that
grows with the system size~\citep{santos2016cooperative}.

\textit{Resonant transition}.--- The periodic driving of $\hat{H}_{0}$
tries to establish transitions between the different bands, but for
this to happen efficiently it must deposit an amount of energy on
par with the gap between the bands, $\omega\approx\Delta_{b}$. For
the initial state considered in Fig.~\ref{Fig_1}, the most relevant
bands are $b=0,1$ and $2$, with the corresponding gaps $\Delta_{1}$
and $\Delta_{2}$. To see the dependence of the heating time on the
dominant gap more clearly, we rescale the driving frequency by the
largest gap, $\Delta_{1}$, as shown in Fig.~\ref{Fig_2}~(b). We
see that the heating time $\tau^{*}$ reaches its smallest value when
$\omega\approx\Delta_{1}$, because at this point we hit a resonant
transition that leads to fast heating. This can be directly observed
also in Fig.~\ref{Fig_1}, where the fastest heating is indeed verified
for $\omega\approx\Delta_{1}\approx11$. Another drop in the value
of $\tau^{*}$ occurs when $\omega\approx\Delta_{2}$, which may be
due to a multiple photon process. Since a square-wave drive contains
multiple harmonics, higher-order transitions might occur. While such
processes, are suppressed at high enough frequencies, at very low
frequencies, they could give rise to a non-monotonic dependence of
the thermalization time on the frequency as is indeed observed in
Fig.~\ref{Fig_2}. Next, we explain what causes the suppression of
heating as the frequency increases above $\Delta_{1}$ and, especially,
when it decreases within the range $\Delta_{2}<\omega<\Delta_{1}$.

\textit{Non-monotonic frequency dependence}.--- The maximum energy
required to flip one spin for \emph{any} initial state, scales like
$J_{\text{eff}}\equiv\sum_{r}r^{-\alpha}\sim L^{1-\alpha}$, with
the system size. For the initial state considered above, $J_{\text{eff}}$
coincides with the largest gap between the energy bands, $J_{\text{eff}}=\Delta_{1}$.
In the high frequency regime, $\omega\gg J_{\text{eff}}$, we expect
slow heating, as indeed observed in Fig.~\ref{Fig_1}~(a,c). For
$\omega<J_{\text{eff}}$ one might have expected fast heating to occur,
however, because $\Delta_{1}=J_{\text{eff}}$, one photon from the
drive is not sufficient to induce a transition from the band $b=1$
of the initial state to a neighboring band and the dynamics gets confined
to the initial band for a long time, leading to heating \emph{suppression}
and the emergence of the prethermal phase in Fig.~\ref{Fig_1}~(a,c){]}.
In this case, \emph{increasing} the frequency, $\omega\to\Delta_{1}$,
\emph{induces} \emph{heating} due to the approach to the resonant
condition. Therefore, heating suppression can be achieved by going
away from the resonant frequency either by increasing {[}Fig.~\ref{Fig_1}~(a,c){]}
or decreasing {[}Fig.~\ref{Fig_1}~(b,d){]} the driving frequency.

\begin{figure}
\includegraphics[width=8.6cm]{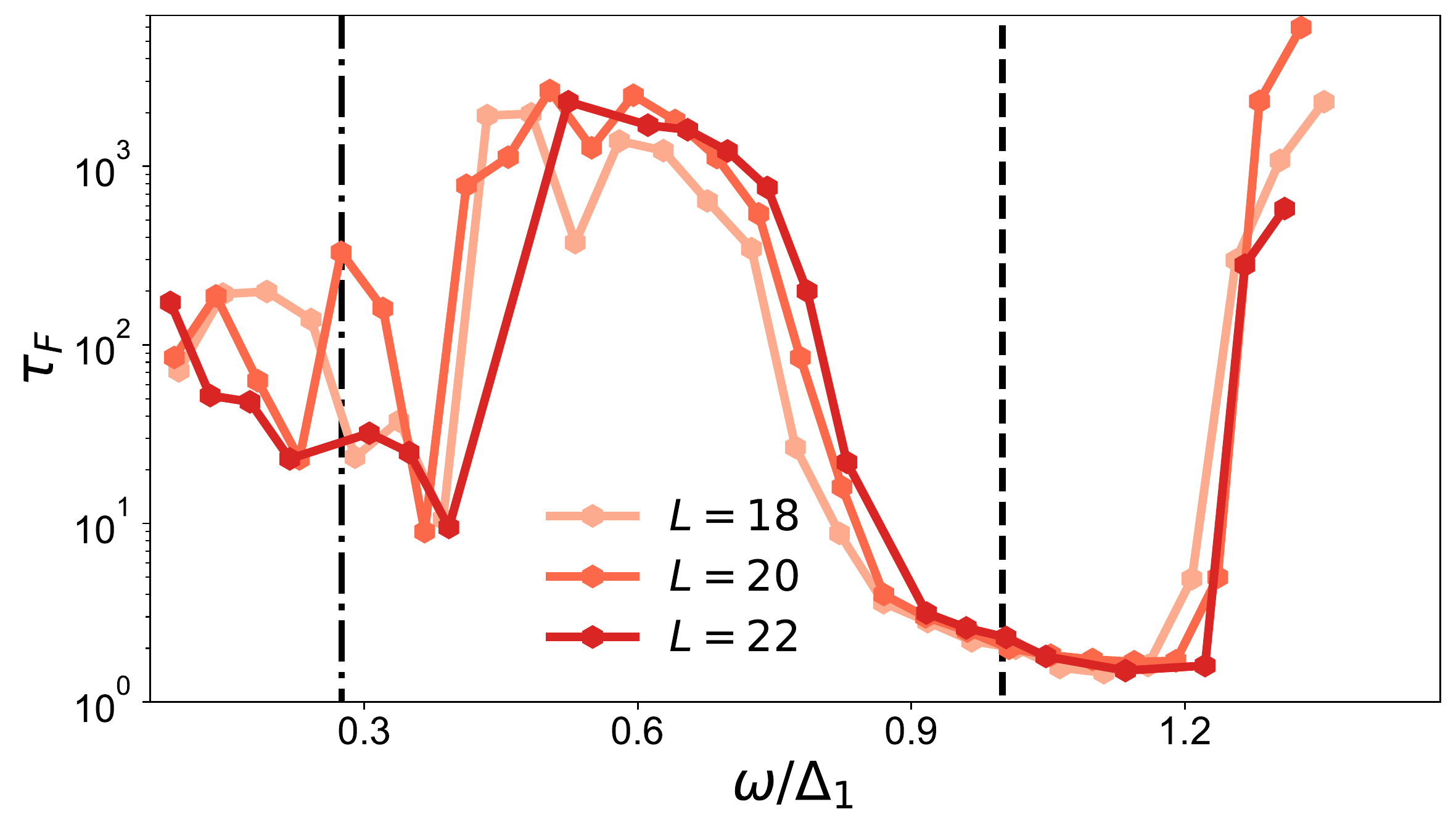} \caption{Half-time decay of the fidelity, $\tau_{F}$, as a function of the
rescaled frequency $\omega/\Delta_{1}$. The two vertical lines indicate
the gaps $\Delta_{2}$ (dashed-dot) and $\Delta_{1}$ (dashed) for
$L=22$. As in Fig.~\ref{Fig_1}: initial state $|\psi(0)\rangle=|11\cdots11011\cdots11\rangle$,
$\alpha=0.67,J_{x}=0.69,h_{x}=0.23,h_{y}=0.21$, and $h_{z}=0.19$.}
\label{Fig_4}
\end{figure}

To demonstrate that for frequencies off-resonance to the gap the dynamics
is indeed confined for long-times to the band of the initial state,
we calculate the fidelity corresponding to the probability to find
the evolved state within the initial band,
\begin{equation}
F_{b}\left(t\right)=\text{Tr}\left[\hat{\rho}\left(t\right)\hat{P}_{b}\right],\qquad\hat{P}_{b}=\sum_{k}\left|V_{k}^{b}\right\rangle \left\langle V_{k}^{b}\right|,
\end{equation}
where $\hat{P}_{b}$ is the projector to the initial band spanned
by the states $\left|V_{k}^{b}\right\rangle $. In Fig.~\ref{Fig_4},
we plot the time $\tau_{F}$ that it takes for the fidelity to decay
to half of its initial value for various frequencies and starting
from an initial state in the band $b=1$. We obtain a behavior very
similar to that for the heating time $\tau^{*}$: the fidelity decays
fast for frequencies close to the gap value, $\omega\approx\Delta_{1}$,
and as we move away from it, $\tau_{F}$ increases significantly.
This corroborates our claim that the suppression of heating and the
emergence of Floquet-prethermalization, that we observe, are indeed
a result of the confinement of the dynamics to the initial band.

\textit{Discussion.---} We demonstrate that in periodically driven
spin systems with long-range interactions, heating can be strongly
suppressed not only with driving frequencies larger than the energy
it costs to flip a single spin, but also with frequencies smaller
than that energy. This is due to the formation of energy bands in
the many-body spectrum of the static system, which get further apart
as the system size increases. If the system is initialized within
one band and the drive is off resonance with the gap between the bands,
then heating is significantly suppressed. This results in a non-monotonic
dependence of the heating time on the frequency. For frequencies larger
than the gap, increasing the frequency suppresses heating, while for
frequencies below the gap, \emph{increasing} the frequencies \emph{enhances}
heating.

Our results therefore provide a robust way to suppress heating even
for small driving frequencies, which can be tested in experiments
with ion traps~\citep{richerme2014nonlocal,kyprianidis2021observation}.
While in this Letter, due to numerical limitations, we have explored
a one-dimensional system, our results should hold for any dimension,
provided $\alpha<d$.

In the future, it would be interesting to see if constraining a long-range
interacting system to a certain energy band allows to obtain, at least
a transient, time-crystalline behavior, which has been ruled out for
$\alpha<d$~\citep{machado2020long}. It would be also interesting
to study the effect of aperiodic drives on heating in such systems~\citep{zhao2021random,mori2021rigorous}.
\begin{acknowledgments}
This research was supported by a grant from the United States-Israel
Binational Foundation (BSF, Grant No. $2019644$), Jerusalem, Israel,
and the United States National Science Foundation (NSF, Grant No.
DMR$-1936006$), and by the Israel Science Foundation (grants No.
527/19 and 218/19). D.S.B acknowledges funding from the Kreitman fellowship.
\end{acknowledgments}

\bibliography{ref}
 
\end{document}